\documentclass[preprint]{article}

\usepackage[nonatbib]{neurips2020}
\usepackage{graphicx}

\usepackage[utf8]{inputenc} 
\usepackage[T1]{fontenc}    
\usepackage{hyperref}       
\usepackage{url}            
\usepackage{booktabs}       
\usepackage{amsfonts}       
\usepackage{nicefrac}       
\usepackage{microtype}      
\usepackage{xcolor}

\newcommand{\Ntrans}{12,159}

\title{Anomaly Detection for Multivariate Time Series of Exotic Supernovae}

\author{%
  V.~Ashley~Villar\\
  Columbia University\\
  New York, NY, USA \\
  \texttt{vav2110@columbia.edu} \\
   \And
   Miles Cranmer \\
   Princeton University\\
   Princeton, NJ, USA \\
   \texttt{mcranmer@princeton.edu} \\
 \And
   Gabriella Contardo \\
Flatiron Institute\\
   New York City, NY, USA \\
   \texttt{gcontardo@flatironinstitute.org} \\
   \And
   Shirley Ho \\
   Flatiron Institute\\
   New York City, NY, USA \\
   \texttt{shirleyho@flatironinstitute.org} \\
   \And
   Joshua Yao-Yu Lin \\
   University of Illinois at Urbana-Champaign\\
   Urbana, IL 61801 \\
   \texttt{yaoyuyl2@illinois.edu} \\
}

\begin{document}

\maketitle

\begin{abstract}
Supernovae mark the explosive deaths of stars and enrich the cosmos with heavy elements. Future telescopes will discover thousands of new supernovae nightly, creating a need to flag astrophysically interesting events rapidly for followup study. Ideally, such an anomaly detection pipeline would be independent of our current knowledge and be sensitive to unexpected phenomena. Here we present an unsupervised method to search for anomalous time series in real time for transient, multivariate, and aperiodic signals. We use a RNN-based variational autoencoder to encode supernova time series and an isolation forest to search for anomalous events in the learned encoded space. We apply this method to a simulated dataset of \Ntrans\ supernovae, successfully discovering anomalous supernovae and objects with catastrophically incorrect redshift measurements. This work is the first anomaly detection pipeline for supernovae which works with real time, online datastreams.

\end{abstract}

\section{Introduction}

Supernovae (SNe) are the explosive deaths of stars which radiate across the electromagnetic spectrum. Astronomers study these events through photometry (broadband imaging) and spectroscopy. SN light is affected by the underlying driving physics and the star's environment; the taxonomy of SNe is based on these factors. Traditionally, SNe are classified through spectroscopy, which is prohibitively expensive for most SNe. However, SNe are \textit{discovered} through wide-field photometric surveys, which capture images of the sky in multiple broadband filters to generate \textit{light curves} of SNe (see Fig. 1). Data-driven methods which flag astrophysically interesting SNe for expensive, multiwavelength follow up are essential as a new observatory, the Vera Rubin Observatory (VRO), comes online in 2022. VRO will discover \textit{millions} of SNe annually, increasing our current discovery rate by 100x. 

Given our limited knowledge of the cosmos, we must prepare for this deluge of data by developing \textit{physics agnostic} algorithms which search for unknown phenomena. In this work, we present an anomaly detection pipeline based on a recurrent autoencoder with no physical priors. We search the learned encoded space for out-of-distribution events and evaluate the success of our algorithm by determining if the anomalous events are indeed of rare physical origin.

In the astrophysics literature, anomaly detection for SN light curves has been explored on archival datasets \cite{ishida2019active, pruzhinskaya2019anomaly} using isolation forests and active anomaly discovery. Convolutional neural nets have been used as a dimensionality reduction technique for gravitational wave time series \cite{2018APS..APRL01027S}. Autoencoders have been used to characterize the variability of quasars \cite{2020arXiv200301241T} and denoising gravitational wave time series \cite{shen2019denoising}. In the broader machine learning literature, several studies have focused on searching for anomalous sections of long time series by extracted features from subsections of the series (e.g., \cite{zhang2019deep, 2019arXiv190104997L}). There has been limited focus on anomaly detection in short, aperiodic, multi-band time series.

\section{Data and Method}

\begin{figure}
  \centering
  \includegraphics[trim={0 5cm 0 0},clip,width=14cm]{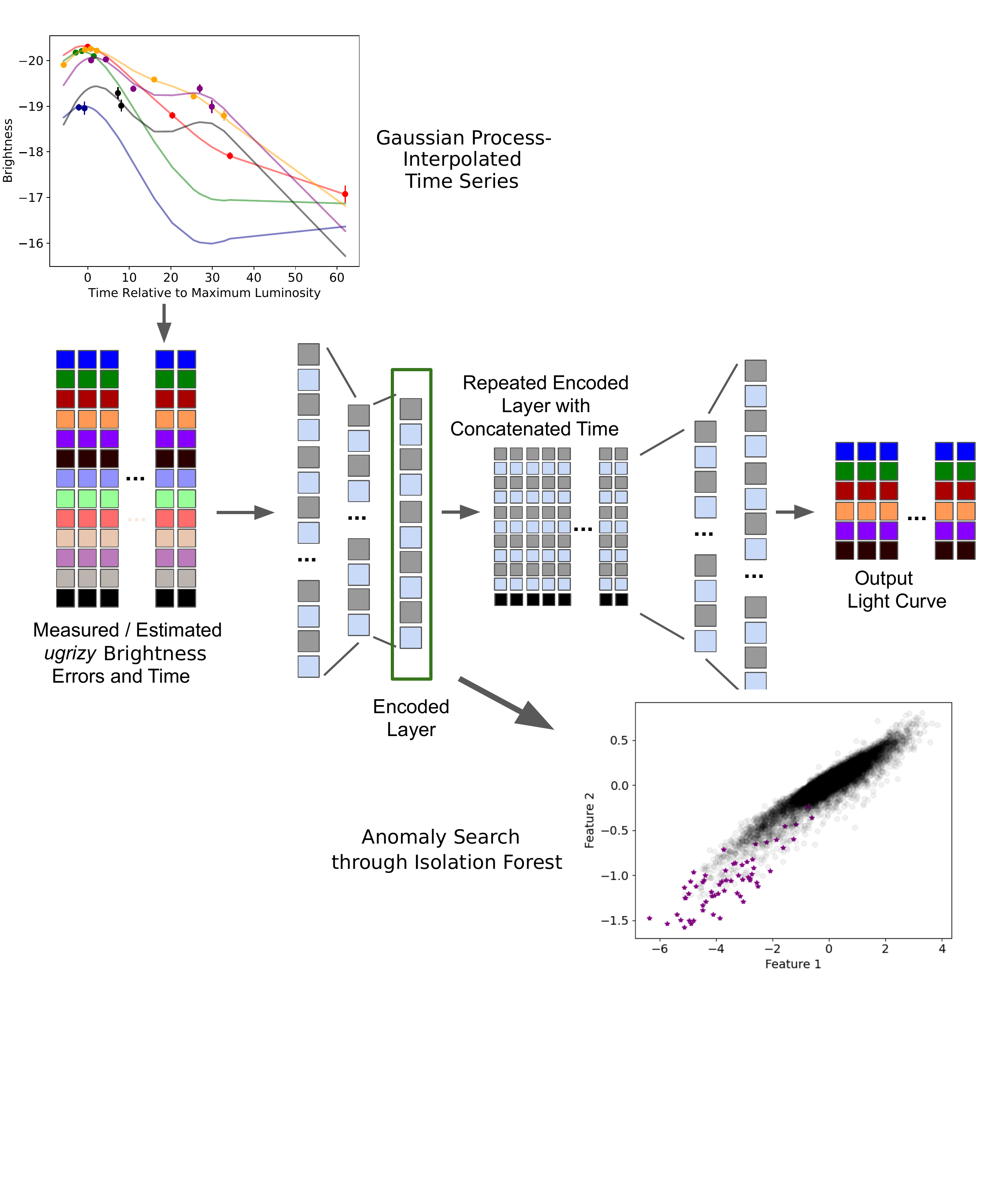}
  \caption{Anomaly detection pipeline, including the recurrent neuron-based variational autoencoder architecture. \textit{Top panel}: The \textit{light curve} (flux versus time as function of filter color) of a SN. The light curve has been interpolated using a 2D Gaussian process. \textit{Middle panel}: The recurrent autoencoder architecture. Light curves are represented as a time series, in which each epoch is represented by 11 features: 6 flux values, 6 estimated error values, and one time. This time series (consisting of $N$ points) is encoded. The encoding layer is repeated $N$ times, each time appended with the associated time value. In principle, this layer allows a user to interpolate and extrapolate light curves by appending time values \textit{not} in the initial time series; however, this feature is not explicitly utilized in this work. Finally, the decoder returns a time series of flux values to be compared to the original time series for training. \textit{Bottom panel}: An example subspace of the encoding space, highlighting anomalous events (as classified by the isolation forest) in purple.}
\end{figure}

PLAsTiCC is a simulation of 3 years of VRO observational data including over 3.5 million transient (including SNe) events from eighteen unique physical classes, extending to redshift $z\approx1.5$ (a measurement of distance). Each event is a light curve made up of observations across six broadband filters ($ugrizY$), as well as metadata about the physics used to generate the light curve. While the PLAsTiCC data was originally used for a Kaggle competition to classify SNe, we re-purpose this dataset as a training set for anomaly detection. Here, anomalous events will be determined by the metadata (i.e., if the event comes from a rare astrophysical origin). In total, our data set contains \Ntrans\ SNe (and SN-like) light curves from twelve extragalactic classes (a random subset of the full PLAsTiCC dataset).  A sample light curve is shown in Fig. 1.  We note that in reality, SN light curves will be contaminated by other galactic astrophysical sources, but it is straight forward to separate extragalactic events from galactic events given a photometric redshift estimate (or more sophisticated methods, e.g., \cite{sanchez2020alert}).

The PLAsTiCC SNe are generated at random distances from Earth following a prescribed volumetric rate. We pre-process the data by correcting the flux using the provided photometric redshift (``photo-z"), including cosmological $k$-corrections. Note that the redshift estimates include realistic error estimates, including catastrophically incorrect estimates; we take these redshift estimates to be correct and discuss how this impact our ability to discover anomalous events in Section 3. We additionally correct the light curves for galactic reddening from dust and cosmological time dilation. The light curves are irregularly sampled in each filter. We use a 2D Gaussian Process to interpolate the data across time to have a flux estimate in each filter at each observed time, with a Gaussian kernel described by:

\begin{equation}
\kappa(t_i t_j f_i f_j; \sigma, l_{t}l_{f})= \sigma^2 \exp\Big[-\frac{(t_i-t_j)^2}{2l_t^2}\Big]\\
\times\exp\Big[-\frac{d(f_i, f_j)^2}{2l_f^2}\Big]
\end{equation}
where $f$ is an integer between 0 and 5 that represents the $ugrizY$ filters, and the parameters $l_t$ and $l_f$ are characteristic correlation length scales in time and filter, respectively.
$d(f_i, f_j)$ is a distance metric between filters. We define it as the Wasserstein-1 distance between each filter's normalized filter, where each filter is treated as a density function in wavelength. Thus, this distance metric is reduced by any overlap in filters, but still simplifies to difference between center wavelengths in the limit of infinitely narrow passbands.
The GP fitting process accounts for the measured data uncertainties, making the light curves more robust to low-confidence outliers.

Astrophysical preprocessing is described in greater detail in \cite{villar2020superraenn}. Importantly, the SN light curves are temporally shifted such that the observed time of peak brightness (in any filter) is considered null time. This means that as new data is taken to form a light curve, the null time is actively updated until the true peak of the SN is reached.

After pre-processing using GPs, our proposed anomaly detection pipeline has two steps: first, we train a recurrent variational autoencoder on the full dataset. Second, we search the learned encoded space for anomalous events using an isolation forest. The variational autoencoder uses recurrent neurons to read in the light curve and estimated GP errors and encodes this light curve as a vector. Before being passed into the decoder, the encoded layer is repeated $N$ times, each time appended with a phase (defined as time since maximum light) $t_i$. The decoder then produces the light curves at the specified $N$ times. This architecture is specifically chosen so that the neural network learns the physical meaning of phase relative to maximum light for a SN. Additionally, the unique repeat layer of our architecture allows us to call for a times not included in the real data, allowing for interpolation and extrapolation of a light curve if desired (although this feature is not used in this work). Figure 1 illustrates our full pipeline, including a schematic of the neural network architecture used.

We then search this encoded space for anomalies using an isolation forest (IF), which has had success in previous studies within the field \cite{pruzhinskaya2019anomaly}. The IF algorithm uses a series of decision trees over random attributes to isolate all events in a sample. Because anomalous events are in some way distinct from the main distribution, the forest requires fewer trees to isolate an anomaly. In this way, each event has an associated anomaly score associated with the path length of the IF. We rank all events by this anomaly score and examine the top $\sim0.5\%$ ($\approx60$) events. 

\section{Results and Discussion}

\begin{figure}
  \centering
  \includegraphics[width=15cm]{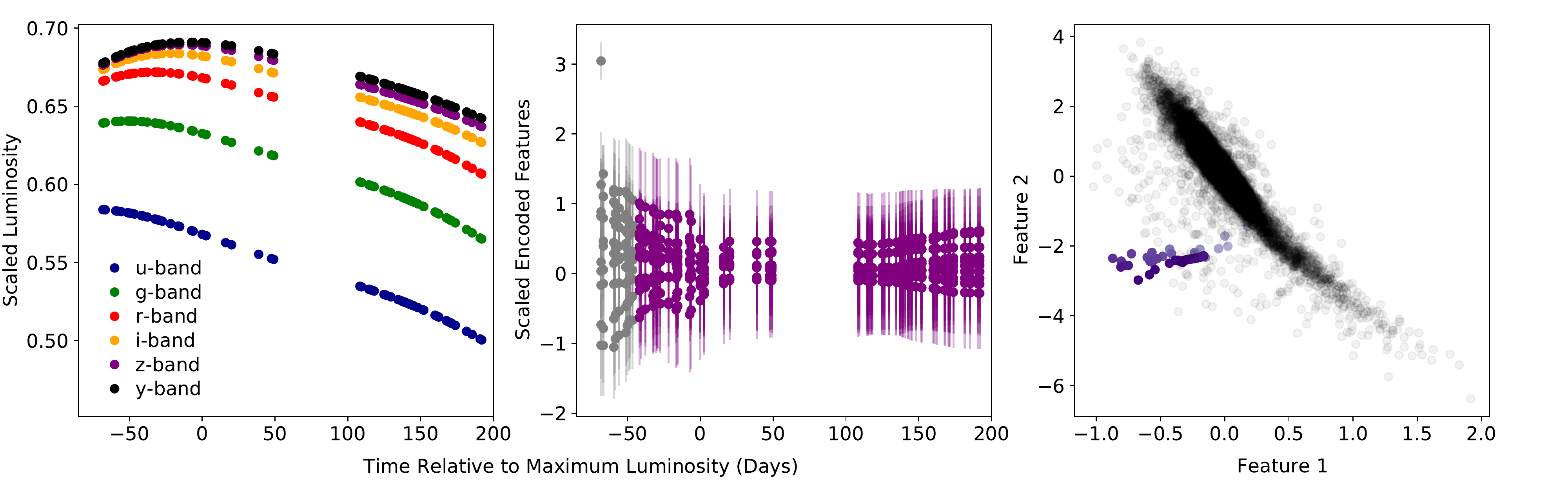}
  \caption{Evolution of an anomalous event detected by our algorithm. Physically, this event is drawn from a rare class of SNe known as Superluminous Supernovae (SLSNe), which would be of scientific interest. For reference, SLSNe make up 2\% of our training set. \textit{Left Panel}: GP-interpolated light curve of the example anomalous event, as a function of time relative to maximum luminosity. \textit{Middle Panel:} Encoded features as a function of time relative to the peak luminosity of the SLSN. About one month before maximum luminosity, the isolation forest correctly identifies the event as anomalous (shown as a switch from grey to purple). Note that the features have been normalized such that their average is equal to zero during the final epoch for visual purposes. \textit{Right Panel:} Evolution of the anomalous SLSN in a subset of encoded space. Black points represent the full sample of events. The encoded values for the anomalous SLSN are shown as squares becoming increasingly purple with time. The SLSN begins in the main distribution of all events (shown in black) and evolves into an anomalous region of feature space. }
\end{figure}

We construct an autoencoder architecture with three recurrent layers of 100 neurons each as the encoder and similarly three recurrent layers as the decoder in Keras \cite{chollet2015keras} with a TensorFlow backend \cite{tensorflow}. We use gated recurrent unit (GRU) neurons with hyperbolic tangent activation functions for most layers, and linear activation functions for the layer immediately before the encoded layer. We train the autoencoder by optimizing a standard loss function which minimizes the weighted mean squared error of the model and Kullback–Leibler (KL) divergence using the Adam optimizer with standard learning rates. We train for 35 iterations. After training is complete, we next train an isolation forest on the encoded light curves using 100 base estimators. 

We rank the final encoded light curves by anomaly score, focusing on the top $\approx0.5\%$ of events (61 events in total). We find that all of these objects are bright, with an average peak magnitude of $-22$ (compared to the sample average peak magnitude of $M_\mathrm{peak}\approx-19$); however, the algorithm is \textit{not} simply selecting the brightest objects in the sample, as many of the brightest objects are not included. Over half of the anomalous events (40 of the 61) are Active Galactic Nuclei (accreting supermassive black holes), which are a rare phenomena (making up $\approx6$\% of the training set) known to have widely varying light curves which are difficult to model and predict. About 20\% (14 events) of the anomalous events are from a rare, bright class of transients known as Superluminous Supernovae (a type of core collapse supernovae from newly born neutron stars). These are correctly considered anomalous events, as they make up just $\approx2\%$ of the training sample and are currently poorly understood astrophysically. Interestingly, the average anomaly score for SLSNe is $\approx2\sigma$ higher than the average score of the sample. The remaining objects (about 20\%) are ``typical" SNe (Type Ia and Type II) which have catastrophically incorrect distance estimates, making the events seem artificially luminous. 

While our algorithm correctly identifies rare transients, it is important that it can do so \textit{before} the transient reach maximum luminosity in order to facilitate multiwavelength follow up. Obtaining spectroscopic and multiwavelength follow-up around maximum light is often necessary to observe early, transient physics within SNe. In Figure 2 we look at one particular example of a SLSN. Our anomaly detection algorithm correctly predicts that this event is anomalous about one month before reaching maximum light. We find that, more generally, our algorithm identifies events as anomalous around maximum light.

Here, we have presented the first anomaly detection pipeline for SNe which works with online data streams. We demonstrated its ability to identify physically interesting and rare SNe during a time window which is scientifically interesting. In future work, we hope to extend this anomaly detection algorithm to data which have not been corrected for redshift in an effort to search for intrinsically dim, anomalous events, as well as include other SN features (e.g., host galaxy information) when evaluating anomalies. Furthermore, we will explore how anomaly score varies as a function of physical class and if this type of network can be used to help classify SNe in real time.

\section*{Broader Impact}

Within the astronomy community, this work will be helpful as a publicly available anomaly detection pipeline to be used when VRO begins observations. We hope this pipeline will promote open science, as we make anomaly scores and the most anomalous events publicly available in real time.

In a broader context, we present an anomaly detection algorithm for multivariate, sparse data in the observed time series is driven from some underlying function correlated across channels. Such a model could be used on similar data sets (such as health- and bio-informatics, climate data, etc.). Because this method is designed to be agnostic to the underlying drivers of the data, this method is not robust to outliers in protected attributes and could therefore be used to inappropriately label anomalies in e.g., social contexts.

\begin{ack}
This work is supported by the Simons Foundation through a Simons Junior Fellowship (\#718240).

\end{ack}

\end{document}